\documentclass[pdflatex,prb,twocolumn,showpacs]{revtex4}
\usepackage{graphicx}
\usepackage{dcolumn}
\usepackage{bm}
\usepackage{amsmath,epsf}
\usepackage[colorlinks,breaklinks,pagebackref]{hyperref}

\begin{document}

\author{Dipta Bhanu Ghosh, Molly De and S. K. De}
\affiliation{Department of Materials Science,\
Indian Association for the Cultivation of Science,\
Jadavpur, Kolkata 700 032, INDIA}

\title{Magneto-optical properties of Europium hexaboride}

\date{\today}

\begin{abstract}
  Full potential spin-polarized self-consistent electronic structure calculations have been performed on LaB$_{6}$, EuB$_{6}$ and YbB$_{6}$ using Linear Muffin-Tin Orbital Method. The electronic structure of EuB$_{6}$ shows a small overlap near the Fermi level. This is in disagreement with the recent ARPES study. The reflectivity spectrum shows a plasma resonance at about 0.3 eV in EuB$_{6}$. Absorption starts at about 1.5 eV and it is due to 4$f$ to 5$d$ transition. Both these features are reflected in the magneto-optical Kerr spectrum. At 0.3 eV, there is a rotation of +8$^{0}$ to -10$^{0}$ and at 1.5 eV, a rotation of about 0.5$^{0}$ occurs. The reasonable agreement with MOKE experiment implies that the inclusion of Drude correction is absolutely essential for the treatment of EuB$_{6}$.
\end{abstract}

\maketitle

\section{INTRODUCTION}

	Divalent hexaborides have been at the focus of attention for the last couple of years because of their anomalous electronic and magnetic properties. Divalent rare earth hexaborides, EuB$_{6}$ and YbB$_{6}$ behave quite differently due to the half and full filled $f$ orbital of Eu and Yb respectively. YbB$_{6}$ is nonmagnetic whereas EuB$_{6}$ orders ferromagnetically\cite{deg97} at 15.1 K accompanied by a large negative magneto resistance and a significant blue shift\cite{brod02} of the reflectivity plasma edge.\cite{deg97} A second phase transition occurs at 12.7 K with a broad peak in the specific heat or an anomaly in the resistivity.\cite{sullow} Neutron diffraction experiment indicates that magnetization increases very sluggishly with decreasing temperature in the absence of magnetic field.\cite{henggeler} The spectacular discovery of ferromagnetism in alkaline earth hexaborides, La-doped CaB$_{6}$, CaB$_{6}$, SrB$_{6}$ and Ba-doped CaB$_{6}$ have aroused deep interest in these compounds.\cite{vonlanthen,ott00,young99} Curie temperature of alkaline hexaborides is above 600 K, much higher than EuB$_{6}$. The introduction of chemical pressure by substituting Ca for Eu in EuB$_{6}$ lowers ferromagnetic transition temperature to 5.5 K.  
\par
	The intriguing magnetic and transport properties of hexaborides arise from the complex electronic properties. Spectroscopic measurements\cite{kimura90} indicate that EuB$_{6}$ may be a narrow gap semiconductor, in unanimity with early\cite{hasegawa80} band structure calculation. However the resistivity measurements revealed metallic behavior below 300 K.[\onlinecite{guy}] Shubnikov-de Haas, de Haas-van Alphen\cite{good} and reflectivity\cite{arko} measurements suggest that EuB$_{6}$ is a semimetal with a very low carrier concentration whereas LaB$_{6}$ is a semimetal with a higher carrier concentration. Model Hamiltonian calculations\cite{cald} claim that electronic properties of EuB$_{6}$ may be explained by considering the fact that it is a semimetal. The anomaly in electrical resistivity suggests that it is close to metal-insulator transition. However the recent optical experiments observe that there is a sizable increase in charge-carrier concentration and/or a reduction of the effective mass of the itinerant charge carriers as it is cooled through the ferromagnetic transition temperature. Hall effect measurements indicate that with the increase in magnetic order, the effective carrier concentration increases significantly.\cite{paschen00} Angle-resolved photoemission and resonant inelastic X-ray scattering experiments reveal that EuB$_{6}$ is a semiconductor with a gap of 0.8 eV.[\onlinecite{denlinger}] Magneto-optical Kerr measurements suggest a Lorentz-Drude model in order to account for a large resonance in its Kerr rotation spectrum.\cite{brod} There is stress upon the importance of the interplay between the charge carriers and the localized-electron interband transitions. All these experiments imply that EuB$_{6}$ has a metallic or a semimetallic ground state. Hence the ground state of EuB$_{6}$ is not only controversial but an open problem.
\par
	First principle band structure calculations based on local density approximation (LDA) yield semimetal with small overlap between the valence and conduction band at the X point of Brillouin zone.\cite{hasegawa79,massidda97} Again these results are within the limitations of density functional theory (DFT). The failure of LDA in describing electronic structure of semiconductors is well known. LDA is also proving to be inadequate for some $s$ and $p$ systems and grossly wrong for the strongly correlated systems like the transition metal and rare earth (RE) compounds. Thus corrections to LDA such as Self-interaction\cite{Svane} and inclusion of on-site Coulomb interaction ( orbital dependent potential ) are required to obtain the exact electronic structure of correlated systems. The band structure calculation performed within weighted density and GW approximations\cite{wu04} has successfully produced a gap in CaB$_{6}$. Strong on-site Coulomb interactions among the 4$f$ electrons in rare earth systems describe their optical properties satisfactorily.\cite{Antonov99,Antonov01,Antonov02,Ghosh} Very recently LDA + $U$ calculations\cite{kunes04} have been performed on EuB$_{6}$ which concludes that its ground state is half metallic semimetal. Kunes et. al. discussed the origin of magnetism in terms of exchange interaction between conduction electrons and local 4$f$ magnetic moments. There have been as many advocates of the semiconducting nature of Eu$^{+2}$B$_{6}$ as of its semimetallic ground state. FLAPW calculations have been reported for EuB$_{6}$ without considering the strong correlations and optical properties of EuB$_{6}$ are not calculated.\cite{massidda97} Magneto-optical spectra provide useful information about interplay between electronic structure and the magnetization of systems. The optical and magneto-optical properties of hexaborides have not yet been studied from the band structure calculation. We have investigated systematically optical and magneto-optical properties of ReB$_{6}$ (Re = La, Eu and Yb). The main purpose of our present work is to observe the effects of $f^{n}$ occupation of Re, where n=0, 7, 14, on optical and magneto-optical spectra and to compare with the experimental results.\cite{caimi04,brod,brod03} 

\section{CALCULATIONAL DETAILS}
Linearized band structure calculations are fairly accurate within the limitations of various approximations used in bypassing the many-body problem. Full potential (FP) self-consistent spin-polarized band structure calculations have been performed on rare earth hexaborides ReB$_{6}$ by the Linear Muffin-Tin Orbital Method (LMTO)\cite{Andersen75} using the FP-LMTO code developed by S. Y. Savrasov et. al..\cite{Savrasov92} Since La has empty 4$f$ shell, the calculations are performed within simple LSDA (local spin density approximation). The DFT is modified in order to include the strong correlations among the $f$ electrons in EuB$_{6}$ and YbB$_{6}$. In the LDA + $U$ method\cite{Anisimov91}, the LDA energy functional is modified by removing the LDA $f-f$ interactions and adding the strong on-site Coulomb interactions among the $f$ electrons. The main idea of the LDA + $U$ method used in this calculation is that the LDA gives a good approximation for the average Coulomb energy of $f-f$ interactions E$_{av}$ as a function of the total number of $f$ electrons. Subtracting this average energy from the LDA total-energy functional and adding orbital- and spin-dependent contributions yield the exact total energy within the mean field approximation. The detailed description of the method is given in Ref. [\onlinecite{Ghosh}].

\begin{table}[h]
\caption{\label{tab:table1}Table contains $a$, the lattice parameter, muffin-tin radius of Re ion, $S_{MT}$, structural parameter $x$, Coulomb parameter $U$ and Density of States at Fermi level, $N(E_{F})$}

\begin{ruledtabular}
\begin{tabular}{cccccccccc}
 &$a$&$S_{MT}$&$x$&$U$&$N(E_{F})$\\
 &$(a.u.)$&$(a. u.)$&&eV&$(St./eV)$\\
\hline
$LaB_{6}$& 7.920 & 3.96 &0.202& 0.0 
& 0.86 \\
$EuB_{6}$& 7.855 & 3.93 &0.202& 9.0 
& 0.36 \\
$YbB_{6}$& 7.832 & 3.92 &0.202& 9.0 
& 0.0 \\

\end{tabular}
\end{ruledtabular}
\end{table}

	The value of the parameter $U = (F^{0})$ used in the present calculation is 9 eV for both the divalent hexaborides in order to place the occupied $f$ states at about 2 eV below Fermi level, ${E_F}$.[\onlinecite{kitamura}] The other Slater integrals $F^{2}, F^{4}$ and $F^{6}$ for $f$ electrons have been used from the experimental paper by Thole et. al.\cite{Thole85} to calculate the matrices $U_{mm'}$ and $J_{mm'}$. The skeletal structure of these cubic hexaborides may be viewed as a CsCl type lattice with RE ion occupying the Cs sites and Cl replaced by $B_{6}$ octahedral cages. Octahedral cages of Boron ions are linked together along the three Cartesian directions. The structure is formally known as CaB$_{6}$ type crystal structure and has a space group $P_{m}4_{m}(O_{h})$. An internal parameter $x$ determines the ratio between inter and intra octahedron B-B distances. Structural parameters $x$ and lattice parameter $a$ are given in Table I. Table I also provides the parameter $U$ and density of states at $E_{F}$. In SrB$_{6}$, a linear relationship between $x$ and energy gap is found. For $x>0.206$, band gap assumes a positive value.\cite{massidda97}. However we use $x$ = 0.202, as reported\cite{massidda97} for EuB$_{6}$. The exchange-correlation potential in the LDA was calculated with Vosko-Wilk-Nussair parametrization. The basis consists of $6s, 5p, 5d, 4f$ states of cation and Boron: $(n)s$ and $(n)p$, where $n$ refers to the principal quantum number. Charge density, density of states and the momentum matrix elements were calculated on a grid of $242$ $\bf{k}$ points in irreducible Brillouin zone and the $\bf{k}$ space integration was performed using the Tetrahedron method. We have treated orbital momentum quantum number, $m_l$ as a parameter. Eu ion is considered to be in its divalent state and calculations were performed with the $4f^7$ configuration.

\section{RESULTS AND DISCUSSION}

	The self-consistent full potential spin-polarized band structure of EuB$_{6}$ is shown in fig. 1. 

\begin{figure}[h]
\includegraphics[scale=0.3]{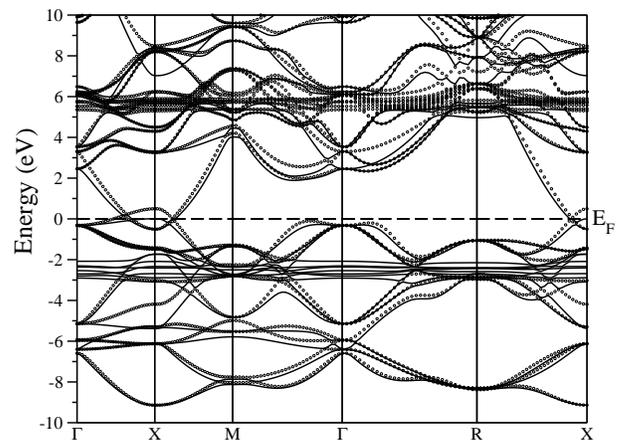}
\caption{Spin-polarized band Structure of EuB$_{6}$.}
\label{fig:eps}
\end{figure}

	Typical small overlap between the 5$d$ conduction band and the Boron $p$ valence bands is seen at the X(100)-point. The lower Hubbard 4$f$ band is situated in between 2 and 3 eV below the Fermi level, $E_{F}$. The upper hubbard band lies about 5.5 eV above $E_{F}$. The bonding 2$p$ orbitals of Boron mainly constitute the valence band near $E_{F}$. The bottom of the conduction band also has some Boron 2$p$ character. Hence the bottom of the conduction band and the top of the valence band is formed from the hybridized Eu 5$d$ and Boron 2$p$ states. This picture is fairly consistent with the XPS and BIS results\cite{takakuwa78} and previous FLAPW calculation\cite{kunes04}. It was suggested by K\"unes et. al. that a ferromagnetic GW calculation may result in a gap opening up in spin-up channel only. However this picture is contrary to most experimental results. On comparison with band structure of LaB$_{6}$, as given in fig. 2, one finds that the Boron 2$p$ bands lie much below $E_{F}$. The Fermi level has Eu $f-d$ hybridized bands located on it. Apparently, this proves that LaB$_{6}$ has much more free carriers than EuB$_{6}$. The unoccupied 5$d$ band is at about 4 eV above $E_{F}$, which is about 1 eV lower than that in EuB$_{6}$. The 4$f$ band lies 2 eV above $E_{F}$ in LaB$_{6}$. The band structure of YbB$_{6}$ reveals a zero gap, with zero overlap at X-point. The 5$d$ states move further up in energy in the conduction band. All the 14 electrons in the 4$f$ state are occupied and the occupied 4$f$ band has a width of 2 eV. The position of the occupied 4$f$ band is 1 eV deeper than that obtained from experiments.\cite{mori93} The blue shift\cite{kimura92} of the 5$d$ band is evident from the movement of the 5$d$ conduction band towards higher energy from LaB$_{6}$ to YbB$_{6}$.

\begin{figure}[h]
\includegraphics[scale=0.5]{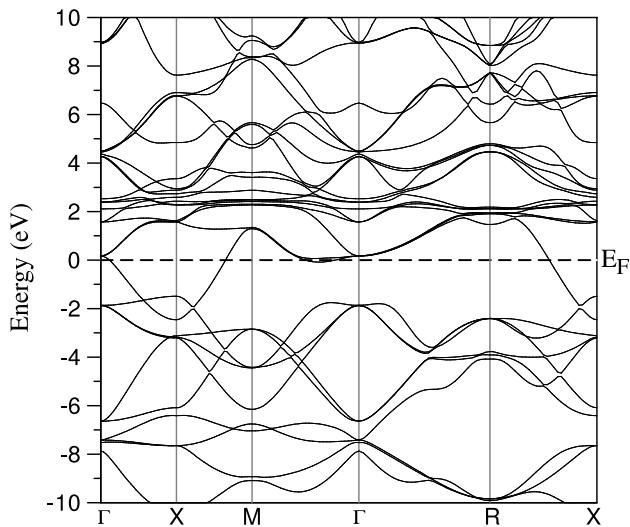}
\caption{Band Structure of LaB$_{6}$.}
\label{fig:eps}
\end{figure}

\begin{figure}[h]
\includegraphics[scale=0.5]{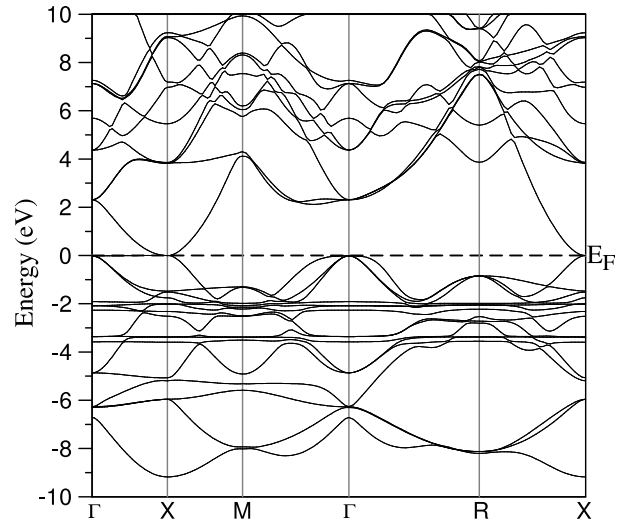}
\caption{Band Structure of YbB$_{6}$.}
\label{fig1}
\end{figure}

\par
The basic difference between the divalent hexaborides and LaB$_{6}$ is that the overlap at the X-point is larger in LaB$_{6}$ and the overlap is between La $d-f$ hybridized and Boron 2$p$ states. Here it is to be kept in mind that the calculation for LaB$_{6}$ is performed within LSDA. The LSDA band structure is in agreement with previous \cite{harima} band structure calculation. In the divalent hexaborides, there is small or no overlap between the Eu (or Yb) 5$d$ and Boron 2$p$ states. Band structure calculation in EuB$_{6}$[\onlinecite{massidda97}] reveal that the electronic structure is sensitive to the dimensions of the Boron octahedra. Choosing a larger ($x >$ 0.206) value of the internal parameter $x$ opens up a small band gap. It is apparent that even this large value is insufficient in producing the gap of about 0.8 eV.[\onlinecite{denlinger}] There is no available band structure calculation on YbB$_{6}$ as far as we know. 

\par
Figure 4 depicts the angular-momentum decomposed density of states (DOS). The region near $E_{F}$ comprises of Boron $p$ and Eu $d$ states in EuB$_{6}$. The fundamental electronic structure of YbB$_{6}$ is very similar to that of EuB$_{6}$ since both the cations are divalent. The occupied $f$ state split into the 4$f^{5/2}$ and 4$f^{7/2}$ states in YbB$_{6}$. The former lies at about 3.5 eV and the latter at about 2 eV below $E_{F}$. The Boron $p$ states lie deeper in LaB$_{6}$ than in the divalent hexaborides.
 
\begin{figure}[h]
\includegraphics[scale=0.6]{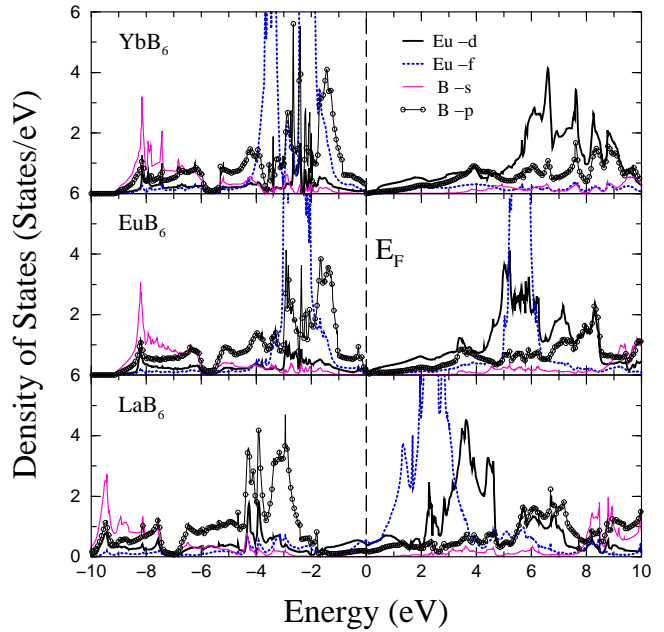}
\caption{$l$-projected Density of States of ReB$_{6}$ (Re = La, Eu, Yb).}
\label{fig:eps}
\end{figure}
The Fermi level is predominantly occupied by the La 5$d$ and 4$f$ electrons. It has no partially filled $f$ orbital. However 5$d$ states are partially occupied, as is evident from fig. 4. That the carrier concentration is very small in the divalent hexaborides is also evident from Table I. The decreasing value of DOS at $E_{F}$ from LaB$_{6}$ to YbB$_{6}$ bears testimony to it. However one cannot compare the two divalent hexaborides directly from this data. The position of the 4$f$ band is fairly consistent with the XPS and BIS results in the divalent hexaborides.

\par 
As mentioned earlier, LaB$_{6}$ and YbB$_{6}$ are nonmagnetic. The spin-polarized calculations yield zero total moment for both of them. EuB$_{6}$ has a total magnetic moment of 6.98 $\mu_{B}$. Orbital polarization is negligible for both the ions. Major contribution to the total moment is from the spin polarization of the 4$f$ electrons. Eu$^{+2}$ has $f^{7}$ configuration. It has completely filled 4$f$ spin-up states and completely empty spin-down states. Hence the spin moment should approach 7 $\mu_{B}$. Orbital moment should be quenched for an $f^{7}$ configuration. Our calculational results are consistent with the atomic analog for Eu. The value of $\mu$ from neutron diffraction experiments\cite{tarascon91} is 7.3 $\pm$ 0.5 $\mu_{B}$. The large error bar comes mainly from uncertainty in the absorption. But the total moment from the present calculation is within that error bar.       

\par
Optical conductivity tensor has been calculated from the standard relations given by Reim et. al..\cite{Reim90} For accuracy calculations have been performed upto 30 eV. The imaginary component of diagonal optical conductivity ($\sigma^{2}_{xx}$) and the real component of off-diagonal conductivity ($\sigma^{1}_{xy}$) have been calculated using the Kramer's-Kronig transformation. A description of the formalism is given in Ref. no. [\onlinecite{Ghosh}]. Optical and magneto-optical spectra have been calculated using the LMTO band structure and further, a Lorentzian broadening of about 0.0068 eV has been used in order to obtain the best agreement with the experimental results in the divalent Re hexaborides. 

\begin{figure}[h]
\includegraphics[scale=0.75]{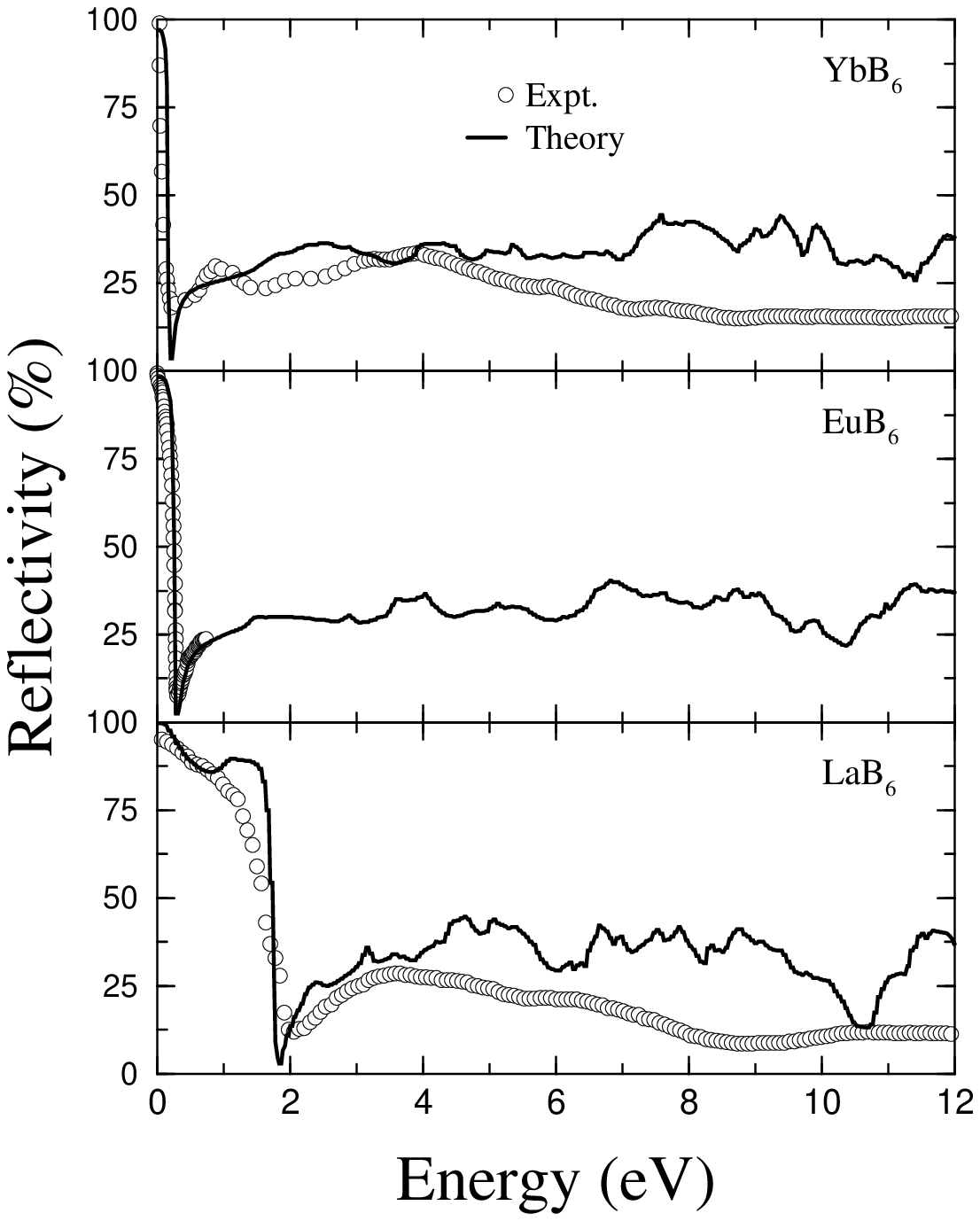}
\caption{Calculated reflectivity spectra of ReB$_{6}$ and their corresponding  experimental(Ref. \onlinecite{kimura90,kimura92,deg97}) reflectivity spectra.}
\label{fig:eps}
\end{figure}

\par
Reflectivity spectra for these compounds are shown in figure 5. The inclusion of Drude correction reproduces the experimental reflectivity spectrum\cite{kimura92} of LaB$_{6}$. An absorption at low energy occurs due to Eu 5$d$ to Eu 4$f$ transition. Due to this transition, a small disagreement occurs at low energy. This can be removed if the 4$f$ states are pushed up by 1 eV. Addition of a constant potential to the Hamiltonian\cite{antonov04} which acts only on the 4$f$ states of La has been successful in reproducing the reflectivity spectrum in La monochalcogenides. A large dip in reflectivity or the plasma edge is successfully reproduced near 2 eV. However for the divalent hexaborides, a very small value of plasma frequency in the Drude correction is required to reach an agreement. The small Drude correction required, howsoever small it may be, strengthens the claim that the divalent Re hexaborides are semimetals with low carrier concentration. Due to smaller number of free carriers, plasma edge effect occurs around 0.3 eV and 0.2 eV in Eu and Yb hexaborides respectively.

\begin{figure}[h]
\includegraphics[scale=0.68]{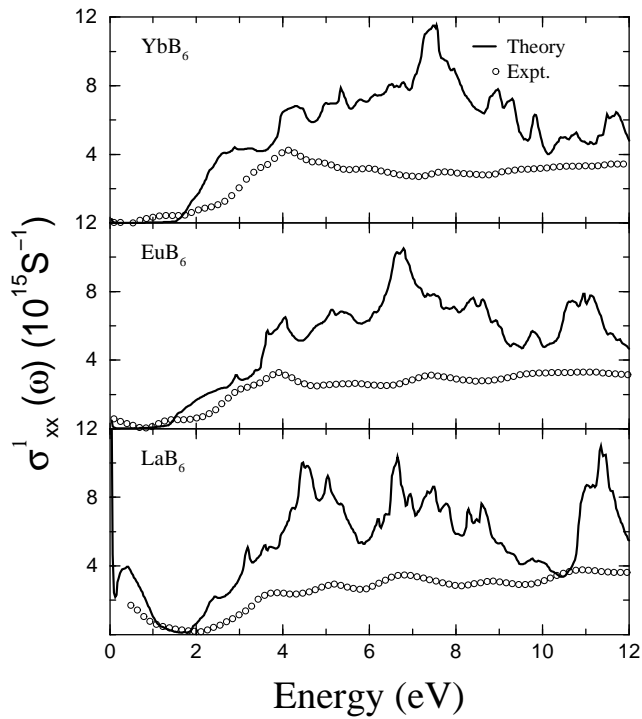}
\caption{Calculated real component of diagonal optical conductivity ReB$_{6}$ and their corresponding experimental (Ref. \onlinecite{kimura92}) counterpart.}
\label{fig:eps}
\end{figure}

\par
The calculated and experimental optical conductivity spectra are depicted in figure 6. Drude correction brings in greater agreement with experiment as far as the plasma minimum is concerned. It occurs at about 2 eV in LaB$_{6}$. But the minima is hardly distinguishable for divalent hexaborides. The broadness of the minimum emphasizes their low carrier density. The first structure in EuB$_{6}$ at 2 eV is due to 4$f$ to 5$d$ transition. This small absorption is missing in the spectrum for LaB$_{6}$ and hence the claim in the last line is justified. This structure shifts very slightly to higher energy in YbB$_{6}$ since the unoccupied 5$d$ states shift to higher energy in the conduction band. The numerous structures till 6 eV are due to transition from Boron 2$p$ state to Eu 5$d$ state. From fig. 4, it is obvious that the largest peak in EuB$_{6}$ will occur at 7 eV and it will be due to transition from the Boron 2$p$ main peak in valence band to the Eu 5$d$ main peak. This peak shifts to lower energy in LaB$_{6}$ and to higher energy in YbB$_{6}$ due to similar blue shift of the 5$d$ state in the conduction band. Around 8 eV, 5$d$ to 4$f$ transition and 2$p$ to 5$d$ transition structures overlap. Beyond 9 eV, the structures arise solely due to transitions from the Boron 2$p$ states to the 5$d$ peaks of Eu.  

\begin{figure}[h]
\includegraphics[scale=0.68]{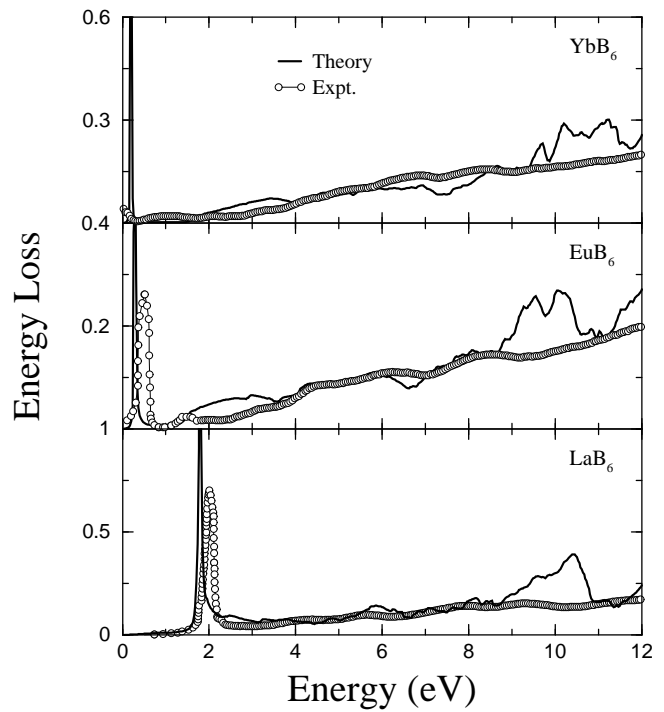}
\caption{Calculated energy loss spectra and the corresponding experimental(Ref. \onlinecite{kimura92}) spectra for the ReB$_{6}$.}
\label{fig:eps}
\end{figure}

\par
The electron energy loss spectrum, as given in figure 7, supports the fact given above. The first peak at 0.2 eV for YbB$_{6}$ and that at 0.3 eV for EuB$_{6}$, are surely due to the plasmon of the conduction electrons. The first peak is found at 1.8 eV in LaB$_{6}$. This reduction in the energy position is due to the fall in the carrier concentration in divalent hexaborides.

\begin{figure*}[t]
\includegraphics[scale=0.9]{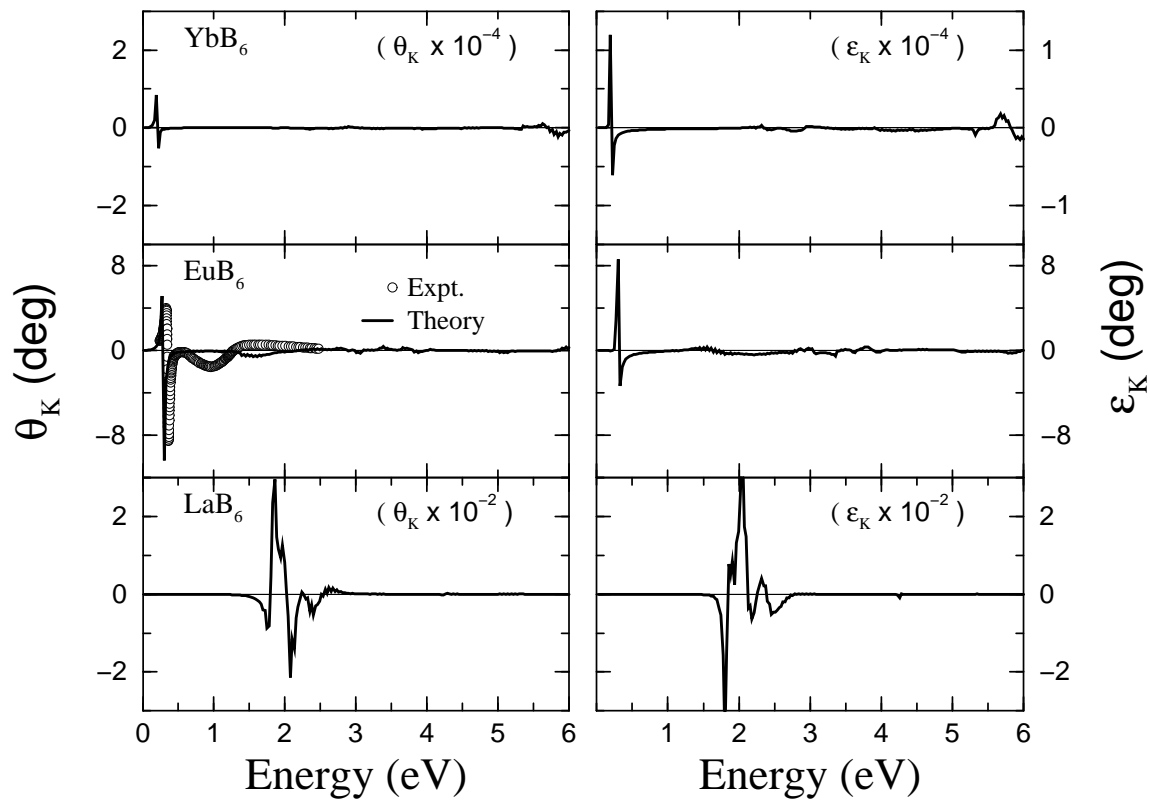}
\caption{Calculated Kerr spectra of LaB$_{6}$, EuB$_{6}$ and YbB$_{6}$ and the experimental Kerr rotation of EuB$_{6}$ at 1.6 K and at 1 T field.(Ref. \onlinecite{brod03}) Calculated Kerr spectra for LaB$_{6}$ and YbB$_{6}$ are magnified by a factor of 2 and 4 respectively.}
\label{fig:eps}
\end{figure*}

\par
Magneto-optical Kerr effect is a very complex phenomenon. Here a very subtle interplay between spin-orbit coupling and spin polarization leads to MO signals. However, in most cases, as in semimetallic magnetic rare-earth compounds, giant signals are obtained due to the plasmon resonance. The most important prescription for obtaining large MO response is a large plasmon resonance followed by a strong interband transition. Inspite of large SO interaction and a fair plasma resonance, we fail to see any kind of MO response in YbB$_{6}$ shown in the left panel of figure 8, as expected for non-magnetic materials. This is also in agreement with experiment.\cite{caimi04} The same happens for LaB$_{6}$. This proves that magnetism plays a very crucial role in MOKE. EuB$_{6}$, with a total moment of 7 $\mu_{B}$, is totally spin polarized. The calculated spectrum, as given in figure 8, shows a jump from $+8^{0}$ to $-10^{0}$ at 0.3 eV and followed by a small signal of about $0.5^{0}$ at 1.5 eV. The former is due to large plasma resonance and the latter is due to 4$f$ to SO split 5$d$ band transition. On comparison with the experimental spectrum, the first signal finds agreement. The second one is slightly suppressed and occurs at 0.5 eV higher in energy. This degree of agreement implies that the Drude correction plays a very important role. Obviously this strengthens the fact that EuB$_{6}$ is semimetallic. Moreover, magnetism plays a crucial role along with SO coupling. 

\par

 Magneto-optical\cite{brod} response in undoped Eu hexaboride has been found to be giant at high fields and low temperature. It has been pointed out by Broderick et. al. that Drude component is essential in reproducing the overall features of the experimental Kerr rotation spectrum. The resonance at 0.3 eV is caused by the response of the itinerant charge carriers to magnetism while the Kerr response at 1 eV is associated with the magneto-optical response of the 4$f$-5$d$ interband transitions. On doping EuB$_{6}$ with isoelectronic Ca, the sharp onset of plasma edge is broadened and the related Kerr resonance is wiped out. This can be due to the large damping of the itinerant charge carriers, which is in turn is due to disorder in the cation sublattice.\cite{caimi04} The intensity of the structure at 1 eV also diminishes with Ca doping since the contribution of the 4$f$ electrons of Eu decreases. These experimental observations are, hence, consistent with the electronic structure calculation of EuB$_{6}$, presented here. 

\section {CONCLUSION}
Self-consistent spin-polarized band structure calculations are presented for LaB$_{6}$, EuB$_{6}$ and YbB$_{6}$. They reveal the metallic nature of the first and the low carrier density nature of the divalent ones. Small overlap at the X-point implies semimetallic nature of EuB$_{6}$ and YbB$_{6}$. This is in disagreement with the recent ARPES measurements in EuB$_{6}$. Present calculations lead to reasonable agreement with optical conductivity, reflectivity and magneto-optical Kerr spectrum. The successful reproduction of the MOKE spectrum in EuB$_{6}$ implies that magnetism, spin-orbit coupling and large plasma resonance are responsible for the large magneto-optical response in this system. The MO response at about 1.5 eV is due to 4$f$ to 5$d$ transition. The inclusion of strong on-site Coulomb repulsion is absolutely essential for its reproduction. Most importantly, Drude-like nature at low energy is a very strong evidence for the semi-metallic nature of EuB$_{6}$ and this is strongly supported by the present calculations. 
 
\acknowledgements
This work is funded by the Department of Science and Technology, Government of India (Project No: SP/S2/M-50/98).

\end{document}